\documentclass[letter]{jpsj2} %% for letters

\title{Heat conduction in one-dimensional lattice dynamical systems\\ far from 
equilibrium}

\author{Akira \textsc{Ueda}$^{1}$\thanks{E-mail: ueda@ms.osakafu-u.ac.jp} 
and Shinji \textsc{Takesue}$^{2}$\thanks{E-mail: takesue@phys.h.kyoto-u.ac.jp}}

\inst{$^{1}$Department of Mathematical Sciences, 
Osaka Prefecture University, Sakai, 599-8531, Japan\\
$^{2}$Department of Physics, Kyoto University,
Yoshida-Nihonmatsu, Sakyo-ku, Kyoto, 606-8501, Japan
}

\abst{We study heat conduction in one dimensional lattice dynamical systems 
far from equilibrium. The Fermi-Pasta-Ulam model and the $\phi^4$ model are
numerically compared to elucidate differences between momentum-conserving and
nonconserving systems. 
As a results, it is found that the heat flux in the $\phi^4$ model does not 
increase monotonically as the temperature differences at the ends of the
lattice is increased, while it does in the FPU chain. 
}

\kword{Fourier's law, far from equilibrium, heat conduction, nanoscale physics, 
one-dimensional lattice
dynamical systems}

\begin{document}
\maketitle

Currently, heat conduction in one-dimensional systems is getting more and
more attention in nanoscale physics.  Murayama studied
heat conduction in the single walled carbon nanotube and 
found that the thermal conductivity behaves like one-dimensional systems.
\cite{Murayama} 
% (5,5) nanotube $B$HEbFM$K$$$o$l$F$b$o$+$i$J$$!#(B
% ``striking'' $B$bJ8L.$NCf$G$7$+0UL#$,$J$$!#(B
Namely, the thermal conductivity a nanotube with a small diameter did not 
converge to a finite value with increase in tube length, but obeyed a 
power law relation as is observed in the one-dimensional systems. 
Similar behavior was observed in the nanoscale world by several 
authers\cite{Yao,Hu0}.

In macroscopic systems, heat conduction obeys the Fourier's law
\begin{equation} 
J=-\kappa\nabla T,
\label{e1}
\end{equation}
where $J$ and $\nabla T$ denote the heat flux and the temperature gradient. 
Recent studies in one-dimensinal systems have clarified that 
size-dependence of the thermal conductivity $\kappa$ is different between
momentum-conserving and nonconserving systems.  For example, the thermal
conductivity diverges as $\kappa\sim N^{0.37}$ in the FPU chain, where
the momentum is a conserved quantity\cite{Lepri1}.  On the other hand,  
it converges in the limit of large system size in the $\phi^4$ model, 
where momentum is not conserved due to the existence of on-site potentials. 
This principle is also applied to the diatomic Toda chain\cite{Hatano}, the
Frenkel-Kontrova model\cite{Hu,Hu2}, the ding-a-ling model\cite{Mimnagh},
the ding-dong model\cite{Sano}, and many others, though the integrable systems
like the harmonic chain or the Toda lattice show another behavior 
$\kappa\sim N$ and there are some exceptions to the 
principle\cite{Giardina,Kurchan}.  Now, this behavior is understood as an 
effect of long-time tail in
the autocorrelation function of the total heat flux\cite{Lepri2}.

On the other hand, in nanoscale devices, the temperature gradient need not be 
small and heat conduction is not necessarily described by the Fouries's law.   
Thus, heat conduction beyond the linear-response regime can be observed in
nanoscale physics.
Accordingly, we study in this paper properties of heat conduction far from
equilibrium, and particularly 
look into differences between momentum-conserving and nonconserving models.

We consider one-dimensional systems with Hamiltonian of the type 
\begin{equation}
H=\sum_{i=1}^{N}\biggl[\frac{p_i^2}{2}+U(q_i)\biggr]+\sum_{i=0}^{N} 
V(q_{i+1}-q_{i}),
\label{Hamiltonian}
\end{equation}
where $q_i$ and $p_i$ denote the displacement and the momentum of the $i$th
particle, $U(q_i)$ the on-site potential and $V(q_{i+1}-q_{i})$ the 
interaction potential between nearest-neighbor particles.  
We use the fixed boundary condition, whch is represented by putting 
$q_{0}=q_{N+1}=0$ in Eq. (\ref{Hamiltonian}).  
We consider the following two models: 
(i) the FPU model is described by 
$V(q)=\frac{1}{2}q^2+\frac{1}{4}q^4$ and $U(q)=0$ and 
(ii) the $\phi^4$ model is decribed by $V(q)=\frac{1}{2}q^2$ and 
$U(q)=\frac{1}{2}q^2+\frac{1}{4}q^4$. 

Moreover, two heat reservoirs are attached to the system, that is, 
particle $1$ is in contact with a heat reservoir at temperature
$T_L$ and particle $N$ is in contact with another heat reservoir at
temperature $T_R$.  In the present study, we use three kinds of heat
reservoirs, the Langevin thermostats, the Nose-Hoover thermostats, and the
thermal wall.  In the Langevin thermostats, random forces and dissipation
terms are introduced into the equations of motion. Namely,
the equations of motion of particles $1$ and $N$ are 
\begin{equation}
\dot{p}_1=-U'(q_1)+V'(q_2-q_1)-V'(q_1)-\gamma_Lp_1+\xi_L(t)\nonumber
\end{equation}
\begin{equation}
\dot{p}_N=-U'(q_N)+V'(-q_N)-V'(q_N-q_{N-1})-\gamma_Rp_N+\xi_R(t)\nonumber,
\end{equation}
where $\gamma_\alpha$ denotes the coupling strength between the particle and
the heat bath and $\xi_L(t)$ and $\xi_R(t)$ are the Gaussian white noise 
with zero mean and variance
\begin{equation}
\langle \xi_{\alpha}(t)\xi_{\beta}(t')\rangle 
=2\gamma_{\alpha}T_{\alpha}\delta_{\alpha\beta}\delta(t-t'),\quad
(\alpha,\beta=R,L).
\end{equation}
 In the case of Nose-Hoover thermostats, deterministic forces play the role of 
heat
reservoir. Namely, the equations of motion for particles 1 and $N$ are 
modified as
\begin{equation}
\dot{p}_1=-U'(q_1)+V'(q_2-q_1)-V'(q_1)-\xi_Lp_1,\quad
\dot{\xi_L}=\frac{1}{\Theta}\biggl(\frac{p_1^2}{T_L}-1\biggr),\nonumber
\end{equation}
\begin{equation}
\dot{p}_N=-U'(q_N)+V'(-q_N)-V'(q_N-q_{N-1})-\xi_Rp_N,\quad
\dot{\xi_R}
=\frac{1}{\Theta}\biggl(\frac{p_N^2}{T_R}-1\biggr),\nonumber
\end{equation}
where $\Theta$ is the thermostat response time.
In the thermal wall, if a particle hits the wall, it is reflected
with a momentum randomly chosen according to the distribution
\begin{equation}
f(p)=\frac{|p|}{T}\exp\biggl(-\frac{p^2}{2T}\biggr),
\end{equation}
where $T$ is the temperature of the wall.

%$B7W;;K!$K$D$$$F>/$7>\$7$/=q$/$3$H!JDI;n$,$G$-$kDxEY$K!K(B
%$B&$(Bt$B$NBg$-$5!"%+%C%W%j%s%0$NBg$-$5!"7W;;%9%-!<%`Ey!9(B
%$B$I$l$@$1$ND9$5$N%7%_%e%l!<%7%g%s$r$7$F!"2?$rB,Dj$7$?$N$+(B
%

In our numerical simulations,
the 4th-order Runge-Kutta and the 6th-order symplectic integrator were
used in the Langevin and the Nose-Hoover cases.  In the cases of the thermal
walls,  
the 4th-order Symplectic Integrator was employed. 
All the simulations were done with a time step $\Delta t=0.01$ and 
the parameters are chosen as $\gamma_L=\gamma_R=0.5$ and $\Theta=0.1$.  
The temperature of particle $i$ is defined as kinetic temperature
$T_i=\langle p_i^2\rangle$ and 
local heat flux through particle $i$ is given by
\begin{equation}
j_i=\frac{1}{2}(p_{i+1}+p_i)F(q_{i+1}-q_i).
\end{equation}
Both the quantities were obtained with averaging over $10^{10}$ time steps 
in the $\phi^4$ model and $10^9$ time steps in the FPU model after a 
nonequilibrium steady state was reached.

\begin{figure}[t]
\begin{center}
    \begin{tabular}{ c c }
\includegraphics[width=6cm]{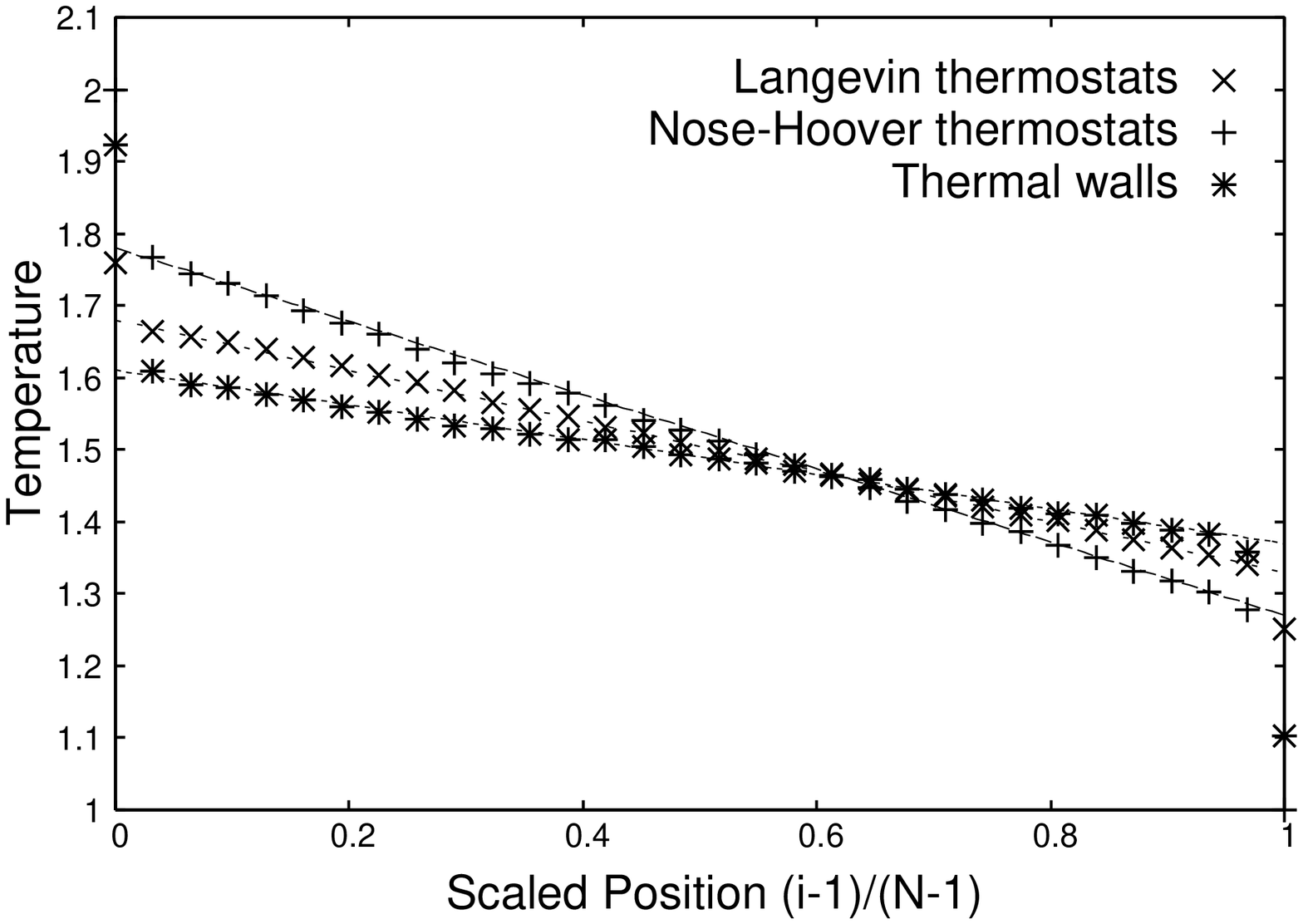} &
\includegraphics[width=6cm]{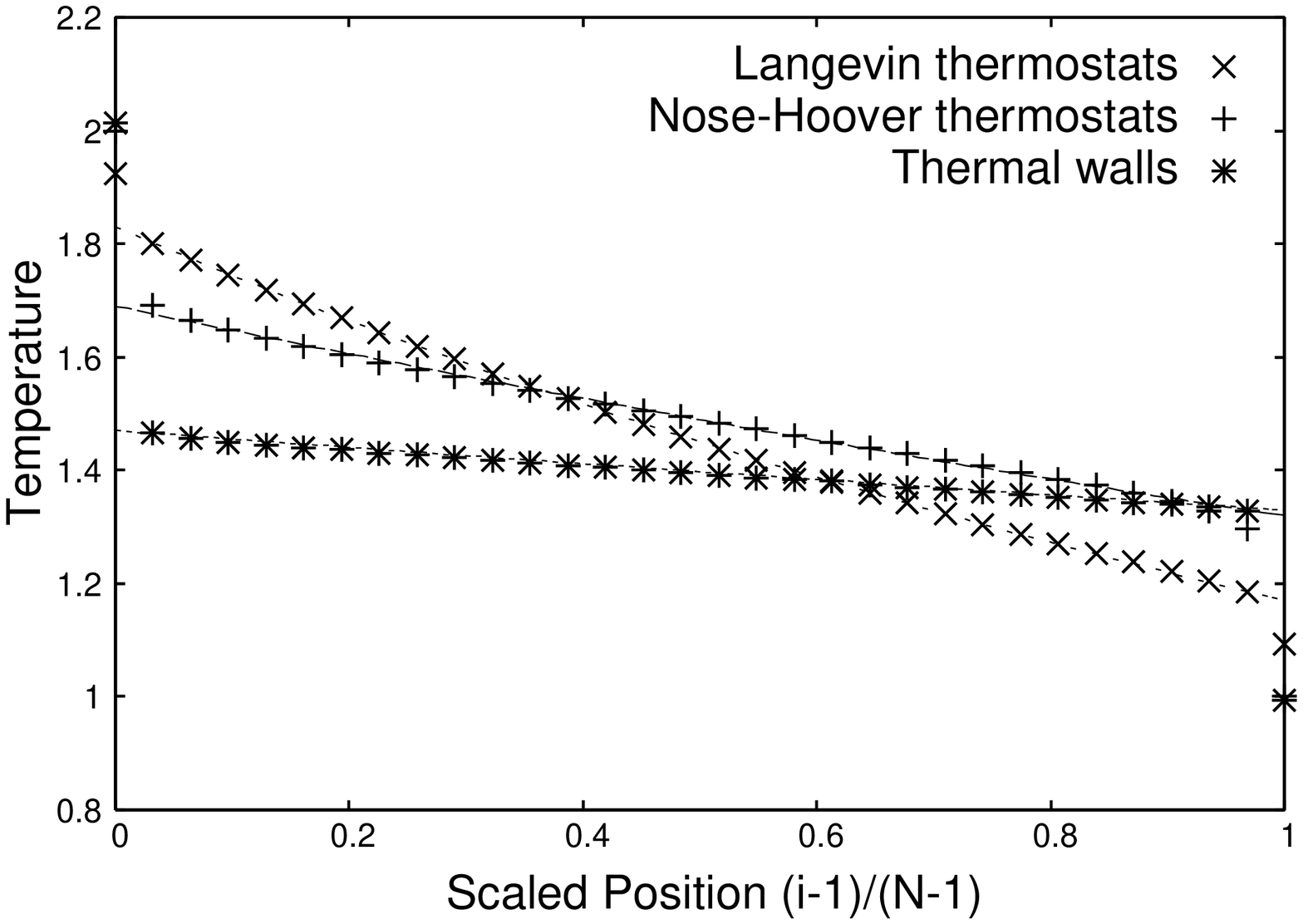} \\
(a) & (b)
    \end{tabular}
\end{center}
    \caption{Temperature profile for the FPU model (a) and the
 $\phi^4$ model (b) with system size N=32 and the Langevein
 thermostats, the Nose-Hoover thermostats and the thermal walls at $T_L=2.0$
 and $T_R=1.0$.}
\label{f1}
\end{figure}

Figure \ref{f1} shows the temperature profile for the $\phi^4$ model and the  
FPU model near equilibrium steady states. 
We see that smooth temperature profile is formed except for boundaries, where 
temperature gaps exist.

\begin{figure}[t]
\begin{center}
    \begin{tabular}{ c c }
\includegraphics[width=6cm]{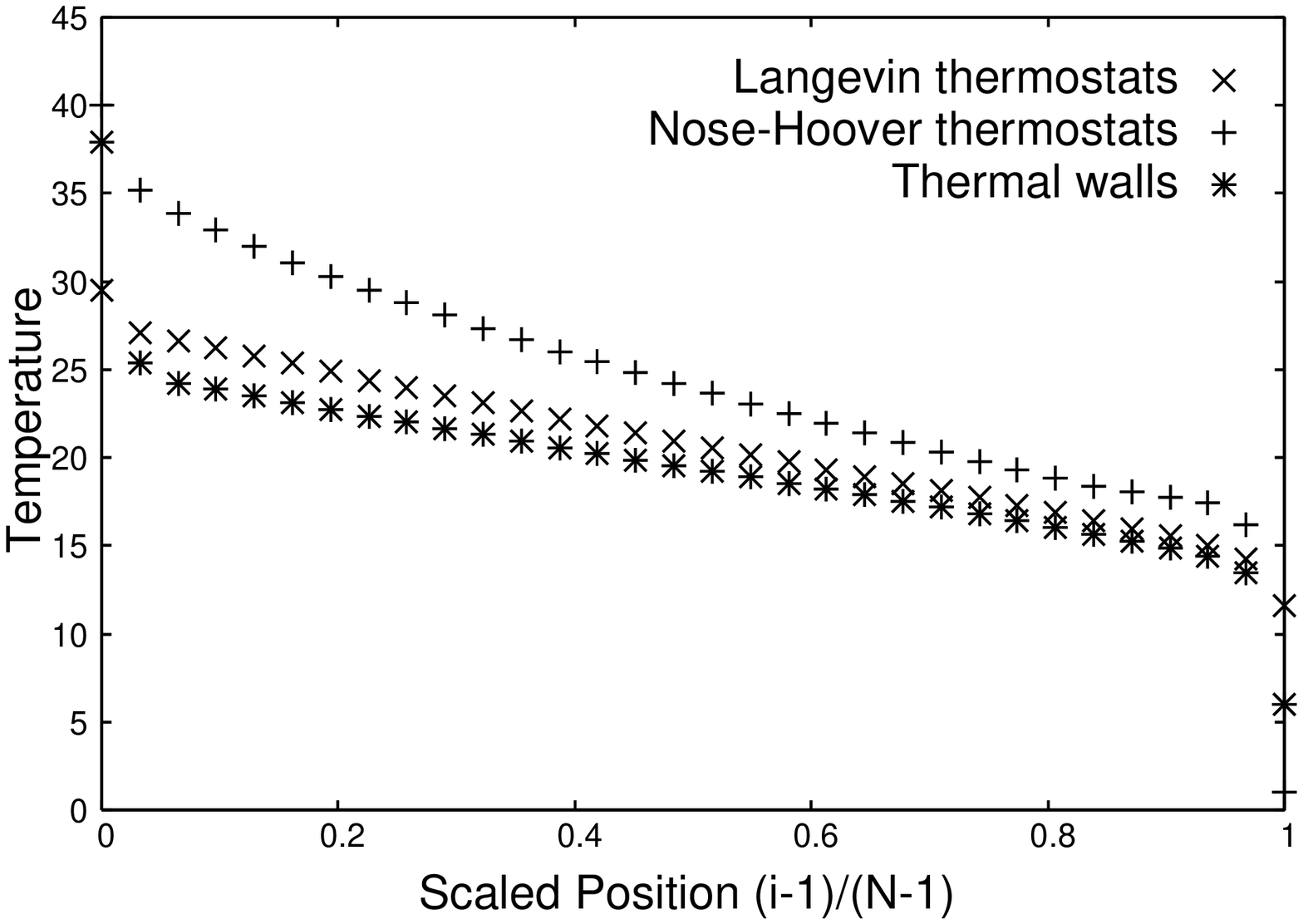} &
\includegraphics[width=6cm]{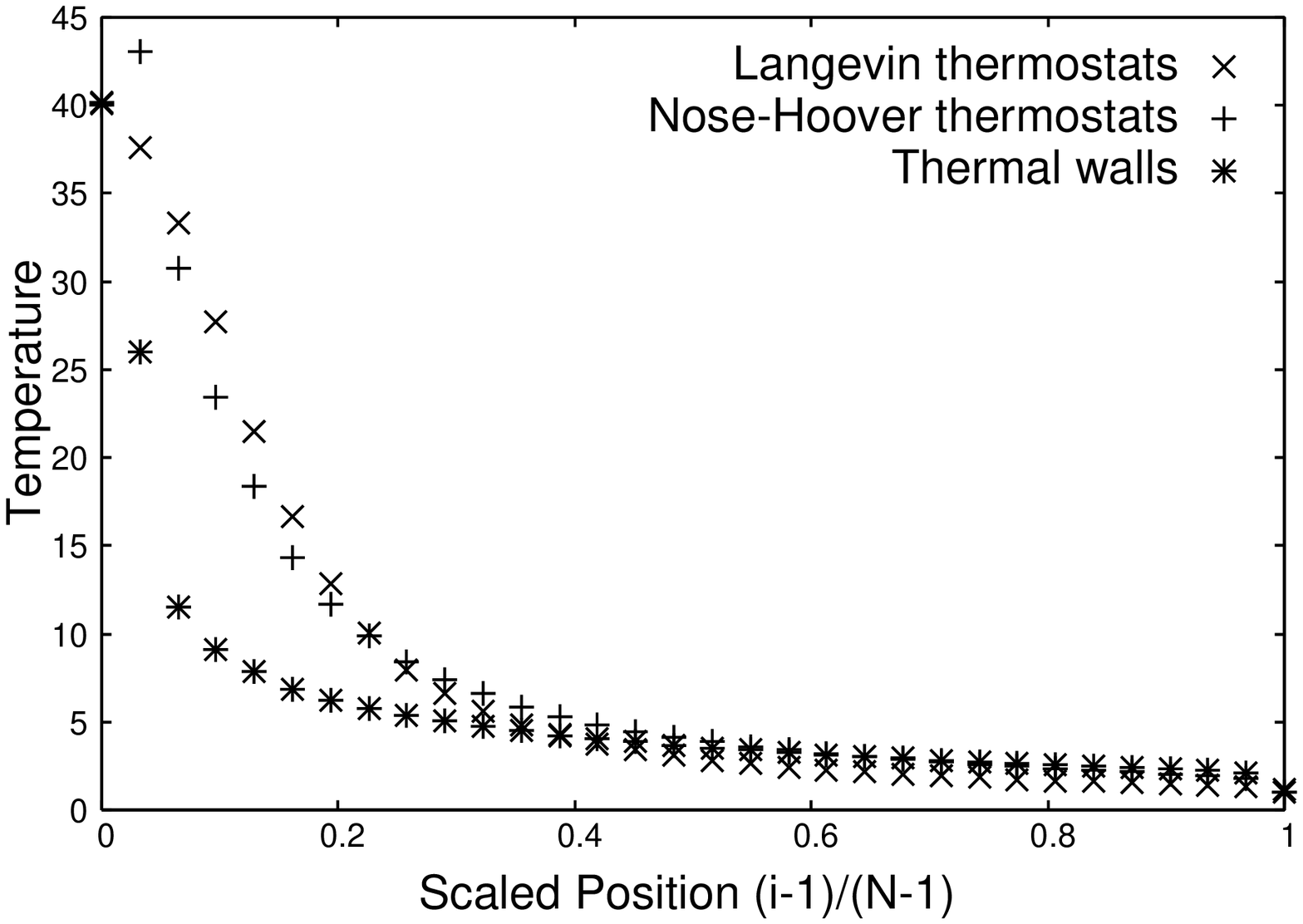} \\
(a) & (b)
    \end{tabular}
\end{center}
    \caption{Temperature profile for the FPU model (a) and the
 $\phi^4$ model (b) with system size N=32 and the Langevein
 thermostats, the Nose-Hoover thermostats and the thermal walls at $T_L=40.0$
 and $T_R=1.0$.}
\label{f2}
\end{figure}

When the temperature difference between the two heat reservoirs is
larger, the system is beyond the linear response regime 
and then Fourier's law does not hold. 
Figure \ref{f2} shows the temperature profile in the $\phi^4$ model and the
FPU model, respectively.  In the FPU model, the temperature profile is
qualitatively similar to the near-equilibrium case.  Only difference is that
the gaps between the heat reservoirs and particle $1$ or $N$ become 
larger than the near-equilibrium case.  On the other hand, in the $\phi^4$ 
model, the temperature profile is drastically transformed.  
Namely, local temperatures are largely shifted to low values in the bulk
region and changes are notable only near the end at the higher
temperature.

\begin{figure}[t]
\begin{center}
    \begin{tabular}{ c c }
\includegraphics[width=6cm]{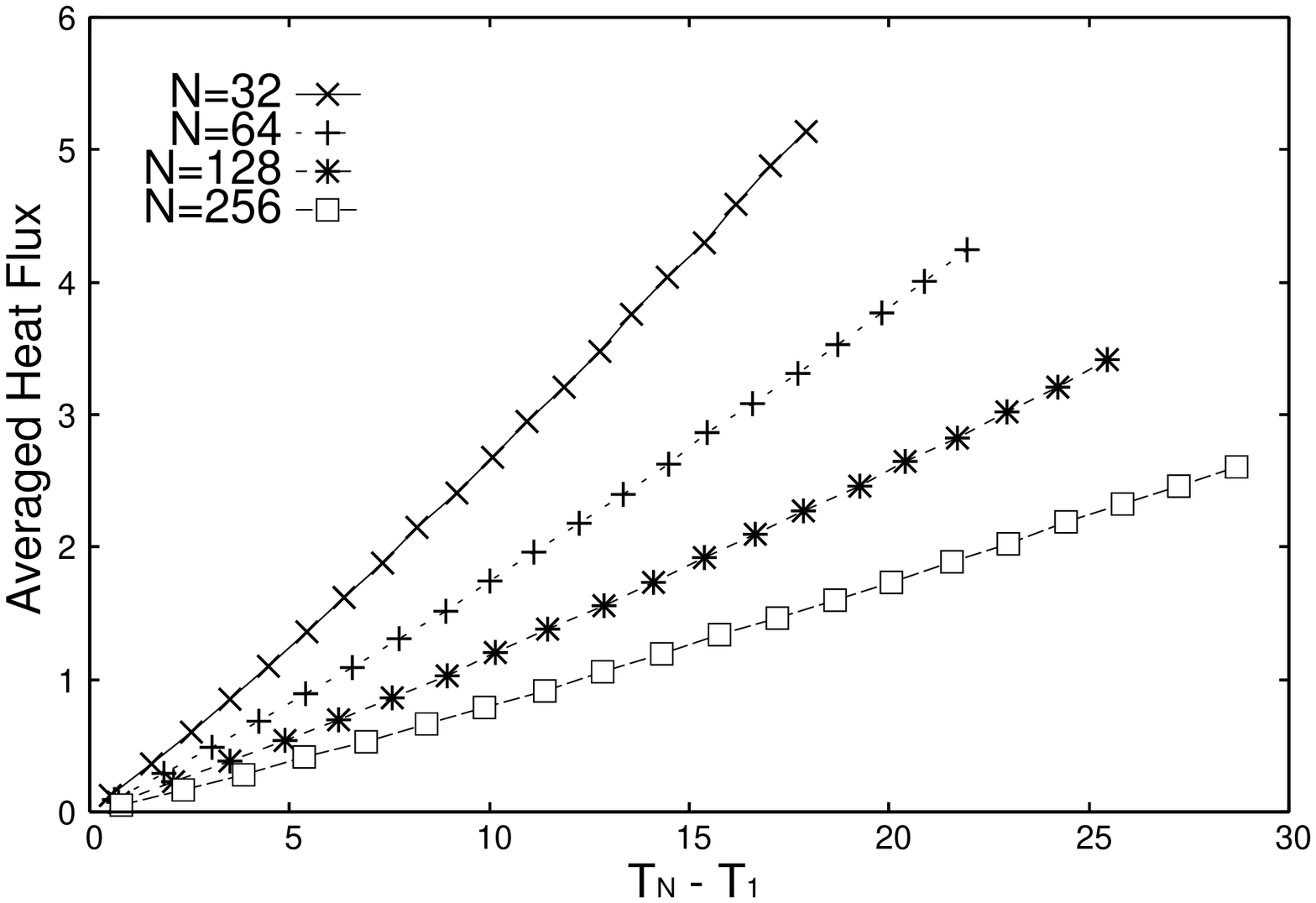} &
\includegraphics[width=6cm]{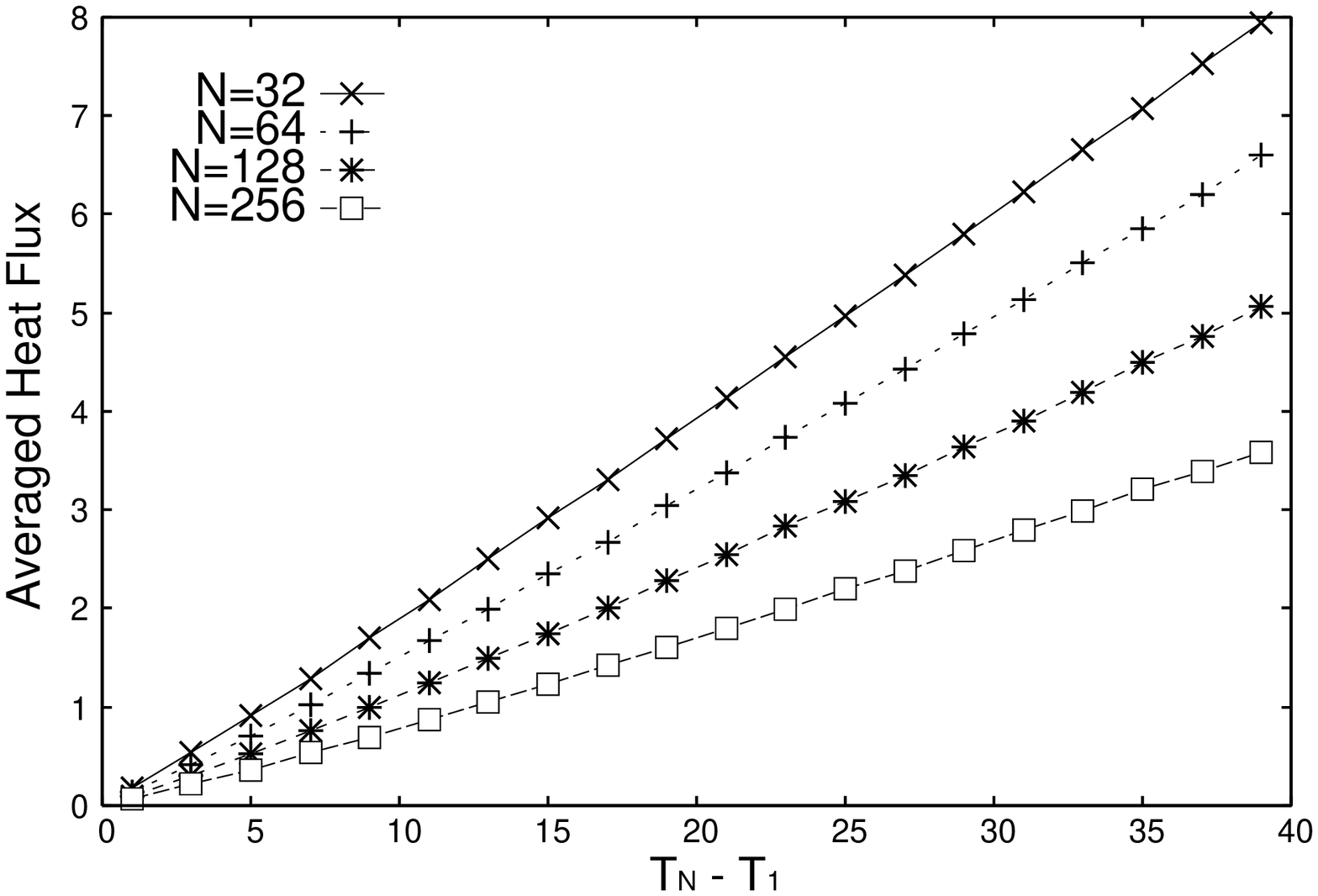} \\
(a) & (b) \\
\includegraphics[width=6cm]{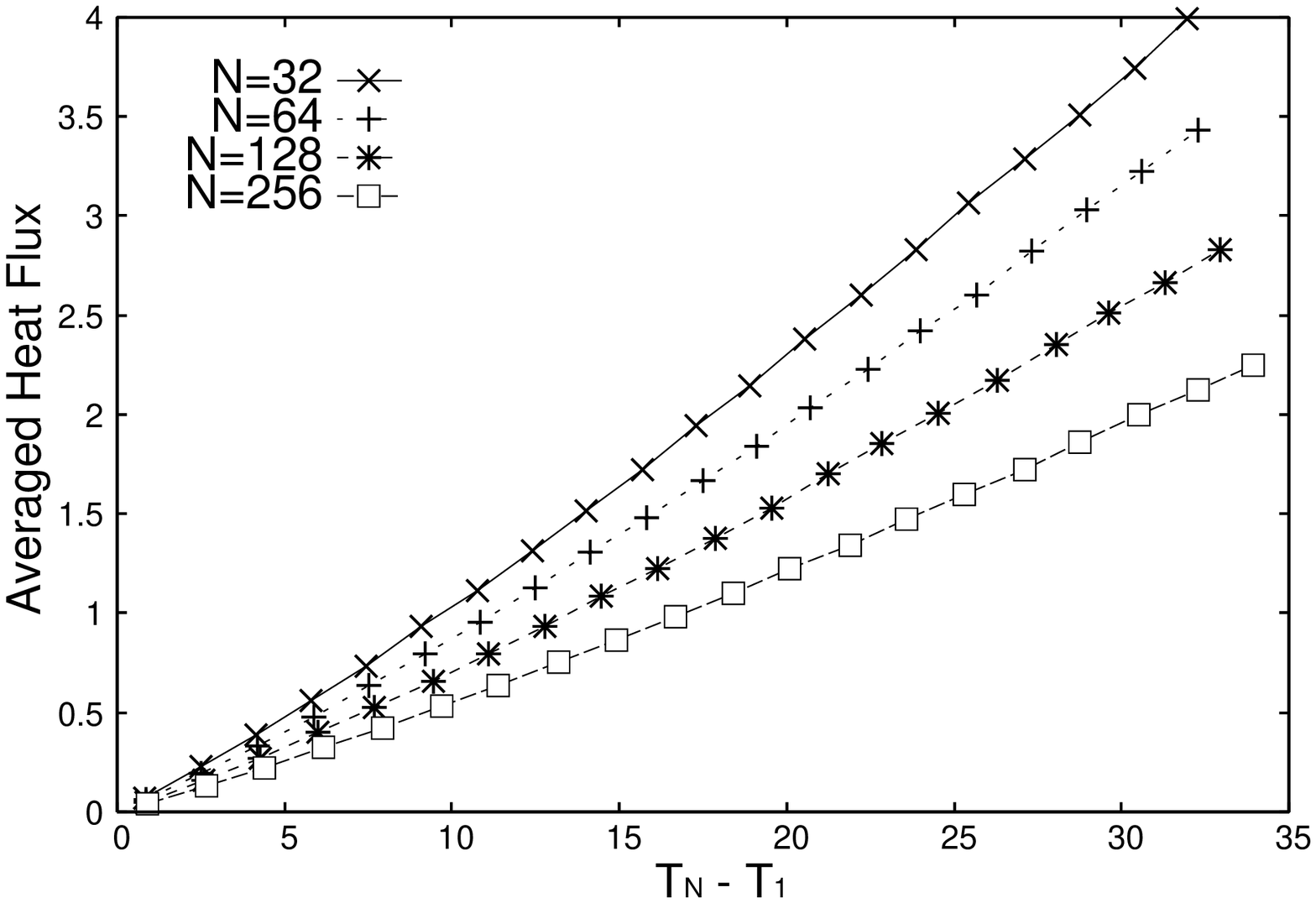} & ~ \\
(c) & ~ \\
    \end{tabular}
\end{center}
\caption{Averaged heat flux and temperature difference at both
 ends for the FPU model with the Langevein
 thermostats (a), the Nose-Hoover thermostats (b) and the 
thermal walls (c).}
\label{f3}
\end{figure}

\begin{figure}[t]
\begin{center}
    \begin{tabular}{ c c }
\includegraphics[width=6cm]{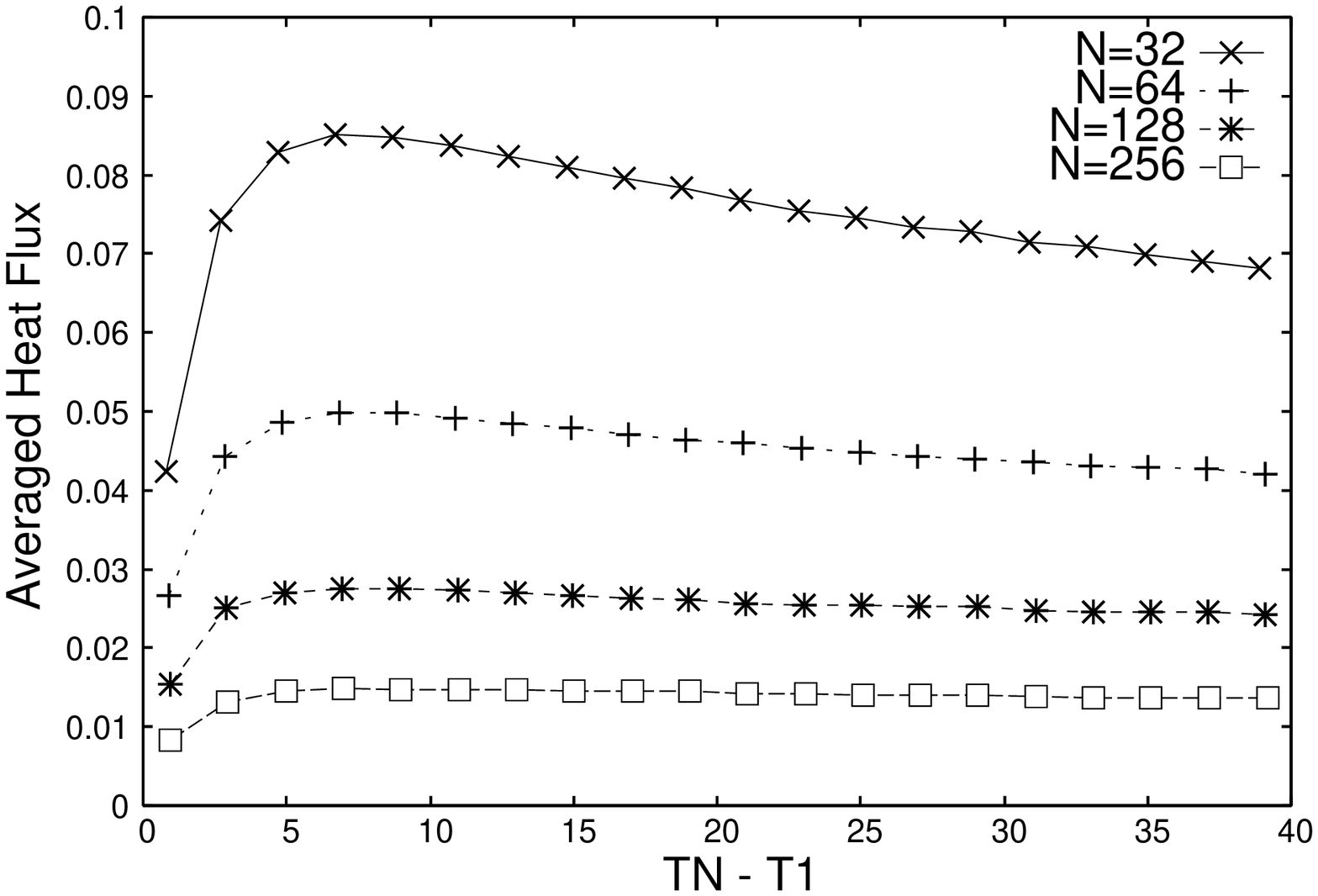} &
\includegraphics[width=6cm]{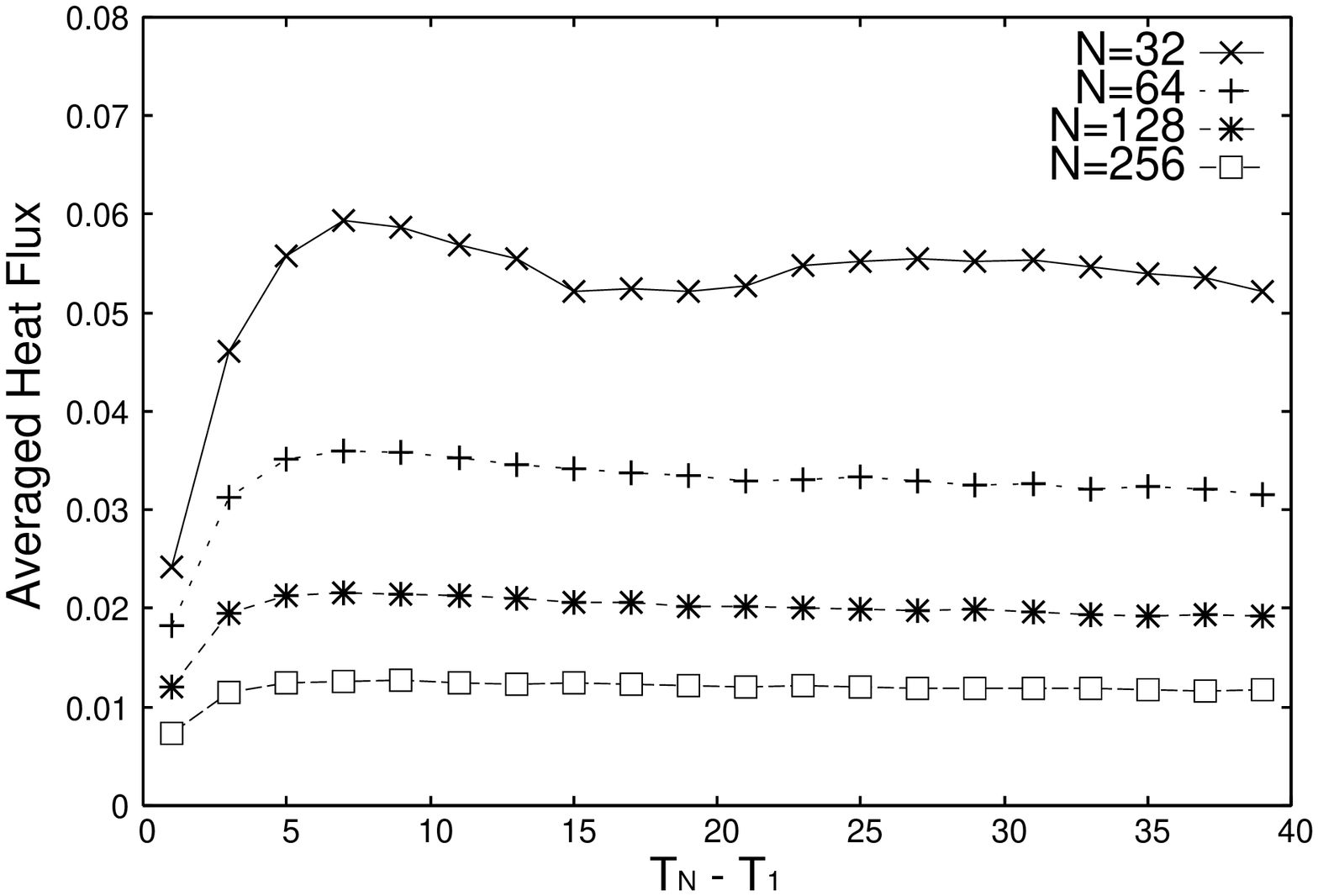} \\
(a) & (b) \\
\includegraphics[width=6cm]{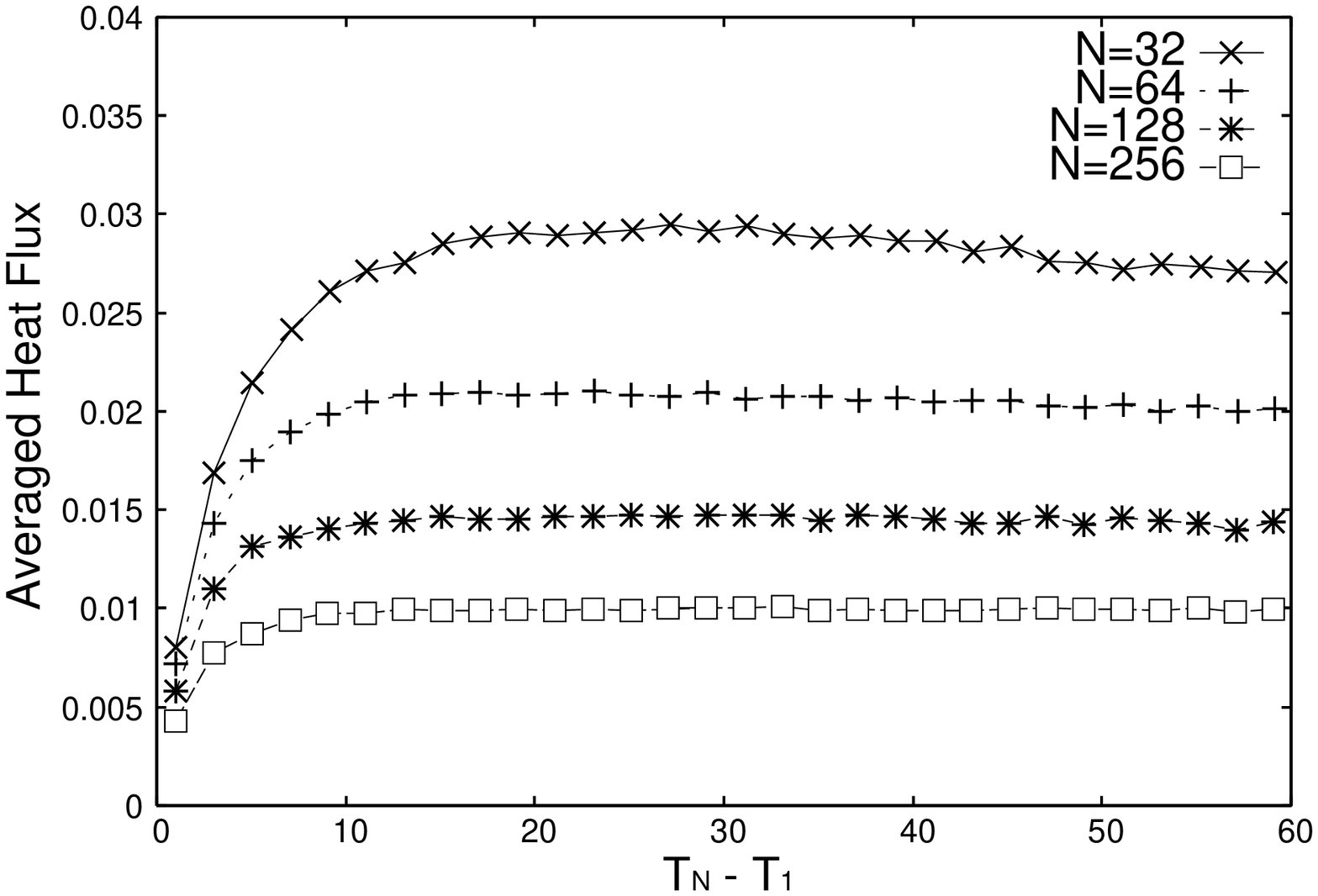} & ~ \\
(c) & ~ \\
    \end{tabular}
\end{center}
    \caption{Averaged heat flux and temperature difference at both
 ends for the $\phi^4$ model with the Langevein
 thermostats (a), the Nose-Hoover thermostats (b) and the 
thermal walls (c).} 
\label{f4}
\end{figure}

Heat flux also shows peculiar behavior.
Figures \ref{f3} and \ref{f4} show the mean heat flux as a function of the
temperature difference $T_N-T_1$. 
\begin{equation}
j=\frac{1}{N}\sum_{i=1}^{N}j_i,
\end{equation}
where $j_i$ is the local heat flux at the $i$th site defined by
\begin{equation}
j_i=\frac{1}{2}(p_{i+1}+p_i)F(q_{i+1}-q_i),
\end{equation}
where $F(x)=-\frac{d}{dx}V(x)$.  It is observed that as the temperature
difference increases, 
the averaged heat flux increases monotonically in the FPU model. 
In the $\phi^4$ model, however,
the averaged heat flux saturates at some value of temperature difference and
does not increase further.
Similar results are obtained irrespective of the types of heat reservoirs.
Therefore, these differences are caused only by the characteristic of the
lattice dynamical systems.  

%$B$3$N7k2L$O2?$r0UL#$9$k$N$+$H$+!#K\Ev$K1?F0NLJ]B8$HHsJ]B8$N:9$H9M$($F$h$$$N$+!)(B
%($B$G$-$l$P!"D4OB3J;R$N%*%s%5%$%HM-$j$HL5$7$H$GF1$87W;;$r$7$F$/$@$5$$!#(B)
%Future problem$B$H$7$F2?$r9M$($F$$$k$H$+!"5DO@$rIU$1B-$9$3$H!#(B

Moreover, we compute the harmonic interaction potential with and without
harmonic on-site potential cases.  In these cases, we confirmed that the 
averaged heat flux increases monotonically.  Thus, the peculiar behavior
of the $\phi^4$ model should be attributed to combination of nonintegrability
and lack of momentum conservation.  Extensive study to clarify the origin of
our findings is planned for the future.
Our results may provide a guideline for designing nanoscale devices.  
 
We have studied the heat conduction far from equilibrium in non
equilibrium steady states in the FPU and $\phi^4$ chains.
As a results, we have found the new interesting phenomenon of the 
heat conduction.  In the FPU model, heat flux increases monotonically as
temperature difference between particles $1$ and $N$.  However, in
the $\phi^4$ model, heat flux does not increase monotonically. 
This property does not depend on the heat reservoirs.

AU thanks H. Nishimori, H. Hayakawa, M. M. Sano for valuable discussions
and helpful comments. This work is supported by the Grant-in-Aid for the
21st Century COE ''Center for Diversity and Universality in Physics''
from the Ministry of Education, Culture, Sports, Science and Technology
(MEXT) of Japan. The numerical computation in this work was carried out at 
Library \& Science Information Center, Osaka Prefecture University.

\end{document}